# NUMERICAL EVALUATION OF THE THERMAL PERFORMANCES OF ROOF-MOUNTED RADIANT BARRIERS

F. Miranville, H. Boyer, F. Lucas, J. Sériacaroupin
Building Physics and Systems Laboratory
40, avenue de Soweto - 97410 Saint-Pierre
La Réunion - FRANCE
frederic.miranville@univ-reunion.fr

**ABSTRACT**
This paper deals with the thermal performances of roof-mounted radiant barriers. Using dynamic simulations of a mathematical model of a whole test cell including a radiant barrier installed between the roof top and the ceiling, the thermal performance of the roof is calculated. The mean method is more particularly used to assess the thermal resistance of the building component and lead to a value which is compared to the one obtained for a mass insulation product such as polyurethane foam. On a further stage, the thermal mathematical model is replaced by a thermo-aeraulic model which is used to evaluate the thermal resistance of the roof as a function of the airflow rate. The results shows a better performance of the roof in this new configuration, which is widely used in practice. Finally, the mathematical relation between the thermal resistance and the airflow rate is proposed.

**KEY WORDS**
Thermal Modelling, Building simulation, Thermal Performance, Thermal resistance.

## 1. Introduction

In hot climates, action must be taken against solar heating to improve thermal comfort in buildings, and to minimise the need for active cooling, which is responsible for heavy electricity consumption. When used widely, particularly in an enclosed context, active cooling results in significant production constraints and also increases pollution, responsible of the global warming.
Solar energy affects the whole building, but especially the roof, because this surface is the most exposed to solar radiation. Therefore, to reduce the energetic contribution from the roof, several solutions are possible, the most often used being thermal insulation.
Mass insulation is generally used and features very low thermal conductivity, of the order of 0.05 $W.m^{-1}.K^{-1}$. Therefore, heat transfer due to conduction is reduced. Mineral wool and expanded foam (polyurethane and polystyrene) are widely known examples.
Nevertheless, this type of thermal insulation does not focus on thermal radiation, which is a non-negligible form of heat transfer, especially in the case of radiating roofs featuring air layers. Devices which help to reduce heat transfer due to radiation are known under the generic name 'Radiant Barrier Systems', or RBS.

## 2. Radiant Barrier Systems (RBS)

### 2.1. Description

The RBS are thin membranes, the faces of which are covered by a material with very low emissivity, which gives them a polished and therefore reflective appearance. Such surfaces release very little thermal radiation compared to surfaces typically used in buildings. They also reflect much of the incident longwave radiation, and hence significantly reduce the radiation heat transfer across the wall into which they have been inserted. Besides, due to an interface composed of low conductivity materials, they also decrease the amount of energy transferred by thermal conduction.

Many products are available, the main differences being the thickness of the RBS (from 1mm to 30mm) and the constitution of the interface. It is to be note that some products feature several reflective layers in the interface and are so called multi-reflective radiant barrier systems.

In France, the usual set-up is to have an RBS inserted between the roof covering and the ceiling, without any other insulation product. The RBS is therefore installed on its own, with air layers being built-in to the roof framework to ensure that radiation heat transfer from the roof covering. Such an arrangement is the cause of coupled heat transfers, which are unlike those which dominate in attics. The general research issue is therefore to develop a methodology to describe the energetic behaviour of such an interface and to assess indicators of its thermal performance.

### 2.2. Literature review

RBS are well documented in the scientific literature and many researchers have contributed to a better understanding of their energetic behaviour. Among the many international publications dealing with RBS, the following subjects were studied in relation with their performances:

- *Assessment of energetic performance*
- *effect of location*
- *effect of the rate and type of ventilation*
- *effect of settling dust on performance*
- *effect of humidity on performance*

Most of these studies come from the United States and for the most part consider an attic, either ventilated or not, featuring an existing nominal thermal insulation (usually mineral wool), characterised by a given thermal resistance R [1] [2]. These studies aim to assess the effect of adding an RBS to the attic, by analysing the factors listed above and the relevant energetic phenomena.

Many experimental and numerical studies have been conducted in order to better put in evidence the thermal performances and the energetic behaviour of RBS installed in attics. Our concern being simulation of walls including RBS, we only focus in the following paragraphs on the numerical studies.

Turning to the theoretical studies, there have been many attempts to model attics incorporating an RBS; nevertheless, these models remain at the macro-volume level, using energy balance equations.

To provide a synthesis, the different attributes of these models are summarised in *Table 1*.

| Models | Fairey | Winiarski | Medina | Moujaes |
|---|---|---|---|---|
| Conduction : Steady state | ✓ | ✓ | | |
| Conduction : Transient state | | | ✓ | ✓ |
| Convection : Mixed, correlations | ✓ | ✓ | ✓ | ✓ |
| Radiation : parallel planes | ✓ | | | |
| Radiation: Radiosity method | | ✓ | ✓ | ✓ |
| Humidity : sorption/disorption of materials | | | ✓ | |
| effect of rafter radiation | | ✓ | | |
| effect of air stratification | | ✓ | ✓ | ✓ |
| effect of air distribution conduit | | | | ✓ |

*Table 1: The different attributes of the attic models*

More detailed models in two or three dimensions, using finite elements or finite volumes, have been developed, but they are rare; most modelling work uses the elements listed in *Table 1*.

Although RBS have been well studied, both from a theoretical and an experimental point of view, the evaluation of the influence of climate variables on their performance was still missing; recently, a study conducted in the United States has led to interesting conclusions in this subject [3]; based on numerical simulations of a standard attic using a transient heat and mass transfer model, several values of performance indicator were obtained for each of the nine defined climates for the United States; in terms of percentage reduction of ceiling heat flux, it was shown that climate parameters having first order effects were local ambient air temperature, humidity, cloud cover index and altitude, while the amount of local solar radiation had no significant influence. Moreover the sample summer integrated percent reduction ranged from 2,3% for mediterranean climate to 38,5% for the humid subtropical one. The used expression for the calculation was the following:

$$Percent\ reduction = \frac{\int_{evaluation\ period} \varphi_{no\_RBS}.dt - \int_{evaluation\ period} \varphi_{RBS}.dt}{\int_{evaluation\ period} \varphi_{no\_RBS}.dt}$$

Where $\varphi_{RBS}$ is the ceiling heat flux when a radiant barrier is present and the ceiling heat flux when no RBS is present in the attic.

## 3. Objectives and methodology

### 3.1. Research issue

RBS are more and more often used in buildings, as insulation products for horizontal and vertical walls; in Reunion Island, characterised by a tropical and humid climate, this technical solution is integrated in roofs, thus creating what can be called complex roofs. These components features an assembling of several materials and the usual set-up is composed of a roof covering made of corrugated iron, a RBS installed between two air layers and a ceiling made of plasterboard. A picture of this complex roof is proposed in *Figure 1*.

As said previously, heat transfer through the whole assembly is a result of combined conduction, convection and radiation. Consequently, assessing the thermal performance of the wall with conventional methods is not appropriate. These methods, the guarded hot box for example, suppose that the product is tested on his own, against heat conduction. As RBS are thin and made of materials with quite high conductivity, results obtained with such methods are not representative of the performance of the complex roof.

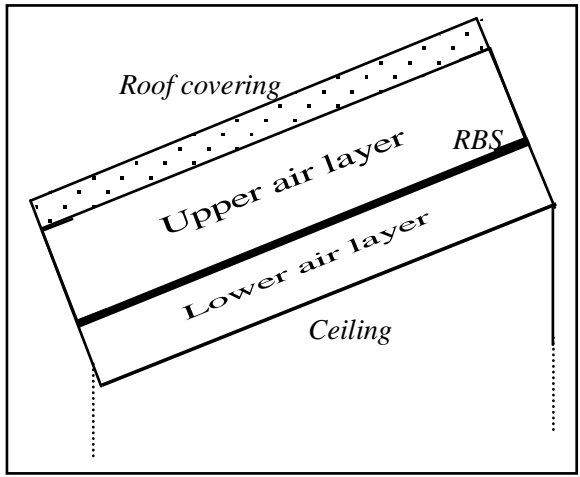

*Figure 1: Section of a standard roof with RBS*

Moreover, the main specificity of a complex roof integrating a RBS and consequently air layers, is the possibility of a ventilation flow rate. In actual configurations, the upper air layer of this type of roofs is naturally ventilated; it is thus important to have information about the effect of the ventilation airflow rate on the thermal performances of the roof.

The aims of the work presented here, based on numerical simulations using a dedicated mathematical model, is thus composed of two parts; on the first hand, the determination of the thermal performances of a standard complex roof including a RBS, and more particularly comparing the results with those of mass insulation, and on the other hand put in evidence the effect of ventilation flow rate on the global performance of the roof.

### 3.2. Methodology

In order to assess the thermal performance of a standard complex roof including a RBS, the proposed methodology is based on numerical simulations. The mathematical model used for the study has been developed especially for parametric simulations of RBS, with a fine coupling with the building. The model is implemented under the Matlab Environment and more particularly in a building simulation code dedicated to the simulations of building insulation components, called ISOLAB. The code integrates for that purpose several models of thermal phenomenon (conduction, convection and radiation) but also a simplified model of aeraulic transfers and a detailed model of humidity transfers in materials.

The proposed methodology is described in *Figure 2*. The several steps include two series of simulations of a standard roof, and in fact, a standard building including a RBS, the first one with a purely thermal model, in order to assess the thermal performance, and the second one with a thermo-aeraulic model, in order to put in evidence the effect of the performance in relation with the ventilation airflow rate.

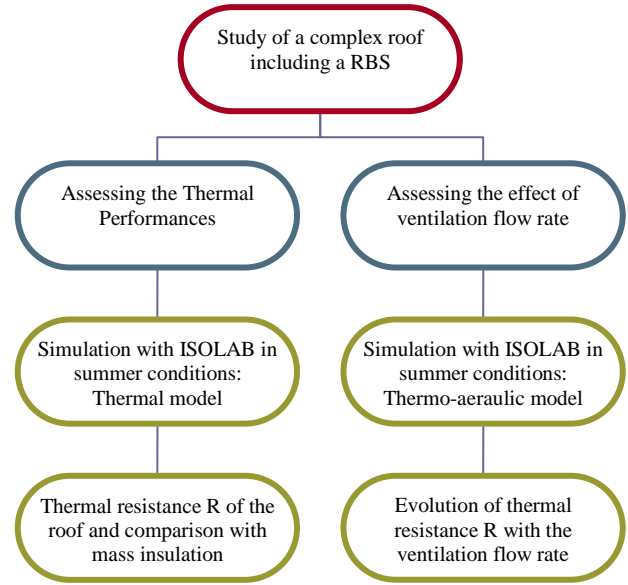

*Figure 2: Methodology of the study*

## 4. Description of the model

### 4.1. The simulation code Isolab

The models used for the simulations of the standard roof were described with the code ISOLAB. This last one, developed under the MATLAB environment, is a prototype of building simulation code, which integrates the hygro-thermal and aeraulic phenomena. However specific developments were done to allow the simulation of complex walls integrating RBS. They concerned the radiation and convection models, and were implemented according to a multimodel approach (the user can choose between several mathematical models for each physical phenomenon). The method of radiosity is thus used for radiative heat transfers and convective heat transfers in the air layers can be described by means of adimensional correlations or of type $h_{ci} = a.(\Delta T)^n + b$.

### 4.2. The Mathematical models

The models used for this study and the building simulation code are fully described in the following papers [4] [5]; the model defined to predict the thermal phenomena occurring in the test cell is composed of 3 macrovolumes or zones as indicated on the *Figure 3*, for which surface temperatures and dry air temperatures are among others calculated. These indicators are useful for the validation of the mathematical model, in particular during the experimental validation when the predictions of the code are compared with the measurements. It is to be noted that the description of the RBS within the model is done through the constitution and the low emissivity of the faces of the separation wall between zones 2 and zone 3. The RBS is indeed described in the model as a wall for which specific thermophysical and radiative properties are used. Besides, although the specific heat of the RBS itself is weak, its inclusion in a roof has to be taken into account by means of dynamic simulations.

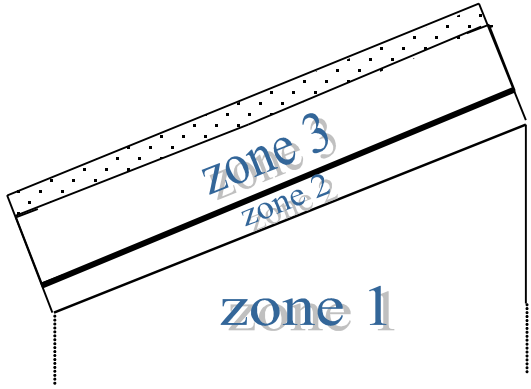

*Figure 3: Modelling of the standard roof*

This thermal model is used for the determination of the thermal resistance of the standard roof. To put in evidence the effect of the ventilation airflow rate on the thermal resistance, the model is completed to take into account the effect of a ventilation flow rate. In fact, through the building simulation code, it is allowed to specify a constant value for the ventilation of a thermal zone; this possibility allow to simulate the ventilation of the upper air layer of the roof. This new version of the model is thus more representative of the actual configuration of a standard roof with RBS and give the possibility to study this advantage of the use of RBS as an insulation product.

# 5. Evaluation of the thermal resistance of the roof

## 5.1. The mean method

As said previously, due to their specificities, obtaining the thermal resistance of complex roofs including RBS with conventional methods leads to non-representative values. Moreover, as we intend to determine realistic values of energetic performance, a method taking into account the dynamic behaviour seems to be more appropriate. The mean Method is thus an interesting tool to assess the desired values. Designed to extract from a database of dynamic physical variables, values of the thermal resistance of walls, this method is used in our case. The method is fully described in the international regulation ISO 9869-1994 [6], and is based on the following formulae, applied on series of dynamic data:

$$R = \frac{\sum_{i=1}^{n}(T_{se,i} - T_{si,i})}{\sum_{i=1}^{n}\varphi_i}$$

Where:
$\begin{cases} R \text{ is the thermal resistance of the wall} \\ T_{si,i} \text{ is the interior surface temperature} \\ T_{se,i} \text{ is the exterior surface temperature} \end{cases}$

Several conditions are necessary to validate the results, and ensure that the energy balance over an entire period is respected.

## 5.2. Application

The mean method is applied to a database of temperature and flux series coming from simulations with the models in summer conditions. In a first step, we focus on the thermal model to calculate the thermal resistance of the roof. Considering a series of temperatures (interior and exterior faces) and flux (through the wall), the result of the calculation using the previous formulae is judge valid in the following conditions:

1. *The percentage of difference ($\varepsilon_1$) between the resistance calculated using the entire series of data and the resistance calculated using the database minus one day is less than 5%*

2. *The percentage of difference ($\varepsilon_2$) between the resistance calculated using the first 2/3 of the series of data and the resistance calculated using the last 2/3 of the database is less than 5%*

Following these criteria and applying the mean method, values calculated are :

$\begin{cases} R = 2,58 m^2.K.W^{-1} \\ \varepsilon_1 = 0.1\% \\ \varepsilon_2 = 2.62\% \end{cases}$

The result is interesting and shows a good performance of the roof under summer conditions; indeed, the calculated thermal resistance is superior as the value obtained for 10cm of glass wool or 6cm of polyurethane foam. It can thus be concluded that the use of RBS is appropriate for standard roofs installed in Reunion Island and that the thermal performance is convenient.

Nevertheless, this result is non-representative of actual physical conditions, because the usual set-up of roofs with RBS features ventilated upper air layers. Our second step is thus the use of the thermo-aeraulic model to determine the effect of the ventilation airflow rate on the thermal resistance of the whole roof.

## 5.3. *Effect of ventilation airflow rate on the performance*

Using the second model, we can focus on the effect of ventilation airflow rate on the thermal performance. Nevertheless, preliminary to the calculation, we have to determine the airflow rate to be use during the simulations. As in actual conditions the wind speed can reach high values, especially in winter in Reunion Island, the calculation should take into account all probable possibilities and therefore cover a large range of possible values. Consequently, a series of 10 values where chosen, ranging from 0 to 4000m$^3$.h$^{-1}$ and thus corresponding to air speed ranging from 0 to 3,6m.s$^{-1}$. After a set of 10 simulations, using the simplified thermo-aeraulic model, considering a constant profile of air speed in the upper air layer, the following results were obtained:

|    | $q_v$ | V | R | $\varepsilon_1$ | $\varepsilon_2$ |
|----|------|------|-------|------|------|
| 1  | 0    | 0.00 | 2.58  | 0.1  | 2.62 |
| 2  | 125  | 0.12 | 4.8   | 0.37 | 1.56 |
| 3  | 250  | 0.23 | 6.77  | 0.76 | 0.67 |
| 4  | 500  | 0.46 | 10.1  | 1.42 | 0.83 |
| 5  | 750  | 0.69 | 12.82 | 1.95 | 2.03 |
| 6  | 1000 | 0.93 | 15.08 | 2.4  | 2.94 |
| 7  | 1250 | 1.16 | 16.98 | 2.75 | 3.72 |
| 8  | 1500 | 1.39 | 18.6  | 3.03 | 4.42 |
| 9  | 2500 | 2.31 | 23.27 | 3.9  | 6.25 |
| 10 | 4000 | 3.70 | 27.35 | 4.68 | 7.68 |

*Table 2: Thermal resistance as a function of the ventilation airflow rate in ther upper air layer*

As we can see, ,the thermal resistance increase with the ventilation airflow rate of the upper air layer, an reaches very high values; indeed, for a ventilation airflow rate of 4000m$^3$.h$^{-1}$, the corresponding thermal resistance is approximately 30m$^2$.K.W$^{-1}$. Even for low value of the ventilation flow rate, the thermal resistance is considerably improved: for example, with 125m$^3$.h$^{-1}$, the performance is quite double of that of 10cm of glass wool.

The curve corresponding to *Table 2* is show in *Figure 4*. The general tendency can be approached by a polynomial regression with the following expression:

$$R = -2.10^{-13}.q_v^4 + 2.10^{-9}.q_v^3 - 7.10^{-6}.q_v^2 + 0.0182.q_v + 2.6155$$

As we can see, the increase of the thermal resistance is an interesting result and show that the thermal performance of the roof can be considerably improved by means of ventilation. This opportunity is relative to roof including RBS, which thus include air layers. The use of the RBS as a major role in the general behaviour of the roof [7]. Physically, the presence of the RBS minimises the effect of radiation and put the emphasis on the effect of convection. Ventilating the upper air layer considerably increase the convection heat transfer in the air layer above the RBS and as a result strongly minimises the heat flux through the insulation product and consequently through the entire wall.

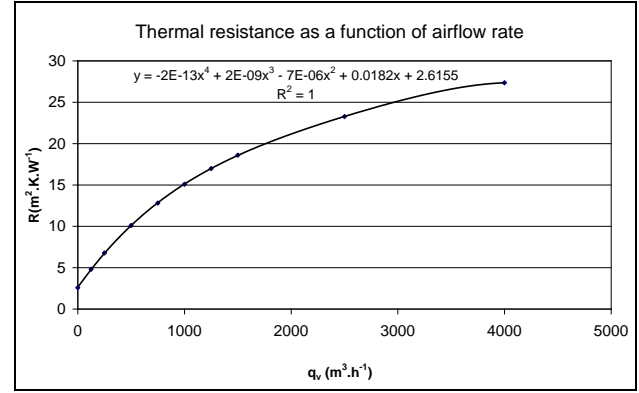

*Figure 4:Thermal resistance as a function of the airflow rate*

## 3. Conclusion

The RBS are more and more used in roofs in Reunion Island and more generally in France. The specificity of these insulation products is their principle of action which consists in a minimisation of the radiation heat transfer through the wall in which they have been inserted.

The research issue for such complex roofs is the determination of the thermal performance, and also the conditions of best performance. This study shows elements of conclusions for both parts of the issue. On the one hand, using dynamic simulations and a method of determination of the thermal resistance of the roof, it is shown that the thermal performance of the roof is convenient, with values superior than the performance of 10cm of glass wool or 6cm of polyurethane foam. On the other hand, the great effect of the ventilation flow rate on the thermal resistance of the roof is put in evidence and show the major advantage of complex walls including RBS [8]. Ventilating the upper air layer lead, even with low values, to a thermal performance considerably better than the non-ventilated case.

Nevertheless, in practice, ventilating the air layer contributes to the accumulation of dust on the RBS, such decreasing its effects on radiation heat transfer. Care should thus be taken to ensure a low ventilation, sufficient for better performances, and with minor effects on the accumulation of dust on the insulation product. The use of air filters may be a possible technical solutions for this problem.

In terms of perspective, the validation of these results according to measured values can be cited. Indeed, all conclusions have been drawn through numeric simulations; comparisons with measurements should put in evidence more complete information.